\documentclass{vldb}
\usepackage{color}
\usepackage{url}
\usepackage{hyperref}
\usepackage{xspace}
\usepackage{pslatex}
\usepackage{enumitem}
\usepackage{epstopdf}

\usepackage{caption}
\usepackage{subcaption}

\usepackage{amsmath}
\usepackage{amsfonts}
\usepackage{euscript}
\usepackage{graphicx}
\usepackage{tabularx}
\usepackage{tabulary}
\usepackage{longtable}
\usepackage{balance}  
\usepackage{microtype}
\usepackage{algorithmicx}
\usepackage{algpseudocode}
\usepackage{comment}

\usepackage[nocompress]{cite}
\usepackage{booktabs}

\usepackage{enumitem}

\usepackage{scalerel}
\usepackage{mathtools}

\DeclareMathOperator*{\argmax}{arg\,max}

\newtheorem{definition}{Definition}
\newtheorem{example}{Example}

\newtheorem{proposition}{Proposition}

\newcommand{\exhaustiveI}{\textsc{\mbox{Exhaustive-I}}\xspace}
\newcommand{\exhaustiveIE}{\textsc{\mbox{Exhaustive-IE}}\xspace}
\newcommand{\greedyI}{\textsc{\mbox{Greedy-I}}\xspace}
\newcommand{\greedyIE}{\textsc{\mbox{Greedy-IE}}\xspace}

\newcommand{\original}{\textsc{Original}\xspace}
\newcommand{\cardinality}{\textsc{CardinalityRelax}\xspace}
\newcommand{\base}{\textsc{BaseRelax}\xspace}
\newcommand{\globalRelax}{\textsc{GlobalRelax}\xspace}
\newcommand{\random}{\textsc{Random}\xspace}
\newcommand{\obj}{\mathcal{F}}

\begin{document}
    
    \makeatletter
    \def\@copyrightspace{\relax}
    \makeatother

\pagestyle{empty}

\title{Improving package recommendations\\through query relaxation}

\numberofauthors{1}
\author{
\alignauthor 
\phantom{${}^{\star}$}Matteo Brucato${}^{\star}$ \hfill
\phantom{${}^{\S}$}Azza Abouzied${}^{\S}$ \hfill \phantom{${}^{\star}$}Alexandra Meliou${}^{\star}$ \\
\vspace{2mm}
\begin{tabular}{cccc}
\affaddr{${}^{\star}$School of Computer Science\phantom{${}^{\star}$}} & & &  
\affaddr{${}^{\S}$Computer Science\phantom{${}^{\S}$}} \\
\affaddr{University of Massachusetts} & & & \affaddr{New York University} \\
\affaddr{Amherst, USA} & & & \affaddr{Abu Dhabi, UAE} \\
\affaddr{\vspace{-3mm}} \\
\affaddr{\eaddfnt{\{matteo,ameli\}@cs.umass.edu}} & & & \affaddr{\eaddfnt{azza@nyu.edu}} \\
\end{tabular}
}

\date{3}
\maketitle

\begin{abstract}
Recommendation systems aim to identify items that are likely to be of interest
to users. In many cases, users are interested in \emph{package recommendations} as collections of items. For example, a dietitian may wish to derive a dietary plan as a
collection of recipes that is nutritionally balanced, and a travel agent may
want to produce a vacation package as a coordinated collection of travel and
hotel reservations. Recent work has explored extending recommendation systems
to support packages of items. These systems need to solve complex
combinatorial problems, enforcing various properties and constraints defined on sets of items.
Introducing constraints on packages makes recommendation queries harder to
evaluate, but also harder to express: Queries that are under-specified produce
too many answers, whereas queries that are over-specified frequently miss
interesting solutions.

In this paper, we study query relaxation techniques that target package
recommendation systems. Our work offers three key insights: First, even when
the original query result is not empty, relaxing constraints can produce preferable solutions. Second, a solution due to relaxation can only be
preferred if it improves some property specified by the query. Third,
relaxation should not treat all constraints as equals: some constraints are
more important to the users than others. Our contributions are threefold: (a)
we define the problem of deriving package recommendations through query
relaxation, (b) we design and experimentally
evaluate heuristics that relax query constraints to derive interesting
packages, and (c) we present a crowd study that evaluates the sensitivity of
real users to different kinds of constraints and demonstrates that query relaxation 
is a powerful tool in diversifying package recommendations. 
\end{abstract}



\section{Introduction} \label{sec:intro}

\looseness -1
The rapid growth of available data has created unique challenges for the field
of database research, and has forced us to rethink multiple aspects of data
management. Most existing work that deals with \emph{volume} in Big Data
targets two directions: increasing data size (more rows) and increasing data
dimensionality (more columns). Along these two directions, database research
has explored time-efficient algorithms that handle expanding data sizes, and
machine learning research has focused on the inference challenges introduced
by the growth in richness and dimensionality of the data. In this paper, we
study a third direction in Big Data that has so far received little attention:
assembling data items into different possible collections of items, or \emph{packages}
(more combinations) \cite{deng, pbq-demo}. From a set of $n$
items, one could identify up to $2^n$ subsets with some desired property;
this makes construction of packages
extremely challenging, even for small datasets. Effectively, even a dataset with
only 100 items can introduce Big Data challenges for a package recommendation
system.

This complexity has deterred data systems from providing
full-fledged support for deriving packages. Yet, the need for package support arises in
many applications.
For example, most airline travel today includes non-direct itineraries that
need to combine several legs, which, as a package, comply with the budget,
schedule, and airline preferences of the traveler.
Package recommendation systems have been used to
derive travel plans~\cite{xie}, team formations~\cite{Lappas2009}, course
combinations~\cite{course-rank,Parameswaran2010}, and nutritionally balanced
meal plans~\cite{pbq-demo}. The items in these packages need to satisfy
criteria for the individual items in the collection, as well as
constraints defined on the entire set of items in the package. 

Introducing constraints on collections of items makes recommendation queries
harder to evaluate: In contrast with traditional
query evaluation, where each individual item is checked against the query
conditions, when evaluating package queries it is not feasible to examine all
possible packages, as the number of combinations grows exponentially~\cite{course-rank,pbq-demo,tran2011evaluation,zhang2011preference}.

Moreover, this complexity poses significant \emph{usability} challenges.
Package queries that are under-specified produce too
many answers, and queries that are over-specified frequently miss interesting
solutions. Consequently, package recommendation systems often produce
unsatisfactory results: 
Would a traveler still prefer her usual carrier if a different airline offered
a travel plan with fewer stops and at a much lower price?
In cases like this, the user often prefers to relax the airline constraint to
improve travel time and airfare.

In this paper, we propose a novel approach to producing package
recommendations that is based on \emph{query relaxation}. In contrast with existing
recommendation systems that search for the top-$k$ packages which satisfy a given
set of constraints, we show that relaxing the package constraints can often
produce preferable solutions. For example, removing the airline constraint
from a travel plan recommendation query may result in a shorter travel time. 
We support our approach with a crowd user study. Our study showed that dissatisfaction 
with the package recommendations that strictly
adhere to user's preferences was very common (30\% of cases). This is
evidence that query relaxation is necessary to provide satisfactory package
recommendations. More stunningly, our relaxation results had overall higher
approval rates (76\%) than the non-relaxed packages (71\%). Our study also
showed that users had different sensitivity to relaxing different types of
constraints.
This indicates that relaxation algorithms can be more effective if they
prioritize constraints based on this sensitivity level.

Our treatment of query relaxation in this context is unique: existing
systems use relaxation to address empty result sets, rather than to derive
alternative recommendations. In summary, our work offers three key insights:
First, relaxing constraints can produce preferable solutions even when 
the original query result is not empty. Second, a solution due to relaxation
can only be preferred if it improves some property specified by the query.
Third, relaxation should not treat all constraints as equals: some
constraints are more important to the users than others. We organize our contributions as follows:
\begin{description}[itemsep=-1mm, leftmargin=0cm]
    \item[Section~\ref{sec:definition}.] We define the problem of deriving
    package recommendations through query relaxation. This is a novel use of
    query relaxation, compared to existing work.

    \item[Section~\ref{sec:heuristics}.] We study the effect of coarse-grained
    relaxations (removal of constraints) to package results. We discuss
    relaxation approaches that aim to improve a package's utility, and
    approaches that balance the improvement in utility with the error due to
    relaxation. We further present greedy heuristics that perform comparably
    to the best-case approaches, and we show that relaxation can be effective
    even when it involves a very small number of constraints.
    \item[Section~\ref{sec:study}.] We present a crowd study that verifies our
    intuition that query relaxation can produce better recommendations and
    demonstrates that users accept relaxations of certain kinds of constraints
    more frequently than others.

    \item[Section~\ref{sec:related}.] We discuss related work in recommendation
    systems and query relaxation. We note that our approach is a novel way of
    improving package recommendation results and differs from the traditional
    use of query relaxation.

    \item[Section~\ref{sec:directions}.] Our approach showed great promise in
    our user study. Our goal is to refine our query relaxation methods to make
    relaxations more targeted and fine-grained; we discuss several directions that we plan
    to pursue towards this goal.
\end{description}

\section[Recommendations via relaxation]{\!\resizebox{0.93\columnwidth}{!}{Recommendations via relaxation}} \label{sec:definition}

In this section, we define the problem of improving package recommendations
using query relaxation. 
A package recommendation query typically involves three types of constraints:
\begin{enumerate}[leftmargin=5mm, itemsep=-1mm]
\item \emph{Base constraints} describe restrictions on individual
items in the package (e.g., each meal should be gluten free, each flight leg should
be no longer than 4 hours, etc). Base constraints are regular selection predicates in
SQL.

\item \emph{Global constraints} describe restrictions on the
package as a whole (e.g., the meal plan should have no more than 1,500
calories in total, the entire trip should cost less than \$2,000, etc). These
correspond to global constraints in CourseRank~\cite{course-rank} and
PackageBuilder~\cite{pbq-demo}, and to compatibility constraints
in~\cite{deng}.

\item \emph{Cardinality constraints} are a special type of global constraints
that restrict the desired number of items in a package.
\end{enumerate}

In this paper, we focus on \emph{conjunctive} package recommendation queries,
i.e., queries that do not contain disjunction or negation of constraints. Some
of the relaxation results that we discuss in this paper generalize to other
cases as well, but we leave a thorough study of 
general query relaxations for
future work.

In addition to
constraints, package recommendation queries also
include
an \emph{objective function}, which is sometimes called \emph{objective criterion},
\emph{objective clause}~\cite{pbq-demo}, \emph{score}~\cite{course-rank}, or
\emph{utility}~\cite{deng}. The objective function describes an attribute that the
package recommendation should either minimize or maximize (e.g.,
minimizing the total airfare, maximizing the amount of protein in a diet, etc).

\begin{example} \label{ex:query}
    A dietitian would like to use a meal planner application to put together a
    nutritious and balanced collection of meals for a client. She would like
    to restrict each meal to no more than 60mg of cholesterol (base
    constraint); she would like each daily plan to include 3 or 4 different
    meals (cardinality constraint) and at least 1,500 calories in total
    (global constraint). To trust that the client will follow the dietary
    plan, she would like to minimize the total preparation time needed
    (objective function).
\end{example}

An answer to the query of Example~\ref{ex:query} is a collection of meals that
satisfies all base and global constraints. Traditional recommendation systems
typically return the \emph{top-$k$} recommendations, which means that out of
all the packages that satisfy the constraints, they would return the $k$
packages with the least preparation time. We challenge this rigid view on
recommendations: If we provide the dietitian with a package where one meal
contains 65mg of cholesterol, but with a drastically lower preparation time, she
might prefer this package to the top-1 non-relaxed result.

We propose to provide \emph{diverse package recommendations} by relaxing the
package constraints. For this paper, we focus on \emph{coarse} relaxations
that remove constraints entirely, rather than changing their
values.\footnote{Several aspects of our work easily generalize to finer
relaxations, but our results show that even coarse relaxations perform well.}
In defining the optimal relaxation, we apply two intuitions: (a) the optimal relaxation should
improve the value of the objective function compared to the top-1 non-relaxed
result, and (b) it should minimize the total deviation from the query
constraints.

\paragraph*{Optimal package relaxation}
We denote a package recommendation query as $Q_{\mathcal{C}, \obj}$, where $\mathcal{C}=\{c_1, \dots, c_n\}$ is a set of base and global constraints, and $\obj$ is an objective function.
Each constraint $c_i$ is a predicate of the form: ${f_{c_i}}~{op}~{\beta_i}$, where $f_{c_i}$ is a function over all the items in a package, $op$ is a comparison operator (e.g., $<$, $\geq$), and $\beta_i$ is a constant in $\mathbb{R}$.  For example, the global constraint in Example~\ref{ex:query} is $sum(calories)\geq 1,500$.
We denote with $f_{c_i}(Q)$ the value of $f_{c_i}$ over the top-1 package for $Q_{\mathcal{C}, \obj}$. 
Similarly, we denote with $\obj(Q)$ the value of the objective function over the top-1 package for $Q_{\mathcal{C}, \obj}$.\looseness -1

\begin{definition}[Package relaxation]
    A package query $Q'_{\mathcal{C}',\obj}$ is a \emph{relaxation} of
    $Q_{\mathcal{C},\obj}^{}$
    if $\mathcal{C}' \subset \mathcal{C}$.
\end{definition}

The top-1 result of a package relaxation may violate one or more
constraints in $\mathcal{C}\setminus\mathcal{C}'$. We use
$Q'\not\models c_i$ to denote that $Q'_{\mathcal{C}',\obj}$ violates
constraint $c_i$. However, as the following proposition shows,
relaxations may improve the value of the objective function:

\begin{proposition}
    Let
    $Q'_{\mathcal{C}',\obj}$ be a \emph{relaxation} of $Q_{\mathcal{C},\obj}^{}$.
    Then:
    \begin{itemize}[noitemsep,leftmargin=0.3cm,topsep=0cm]
        \item If $\obj$ is a maximization objective, then $\obj(Q')\geq \obj(Q)$.
        \item If $\obj$ is a minimization objective, then $\obj(Q')\leq \obj(Q)$.
    \end{itemize}
\end{proposition}

We define the \emph{optimal relaxation} of a package query as the
relaxation $Q^*_{\mathcal{C}^*, \obj}$ that maximizes the
\emph{improvement} in the objective function, while minimizing the
deviation from the constraints of the original query.
More formally:

\begin{definition}[Optimal relaxation] \label{def:best-relax}
The \emph{optimal~relaxation} for a package query $Q_{\mathcal{C}, \obj}^{}$ is a query $Q^*_{\mathcal{C}^*, \obj}$ such that:
\begin{equation} \label{eq:optimal-relax}
Q^*_{\mathcal{C}^*, \obj} = \argmax_{Q_{\mathcal{C}',\obj}^{\prime}, \mathcal{C}' \subset \mathcal{C}}~ \frac{1 + \mathcal{I}(Q';Q)}{1 + \mathcal{E}(Q',\mathcal{C}\setminus\mathcal{C}')}
\end{equation}
\end{definition}

The functions $\mathcal{I}$ and $\mathcal{E}$ are distance metrics that measure the change in the value of the objective function and the amount of constraint violation of the relaxed query, respectively.
In our implementation and experiments, we modeled $\mathcal{I}$ as the \emph{percentage improvement}
in the objective function: $\mathcal{I}(Q';Q) = \frac{|\obj(Q) - \obj(Q')|}{\obj(Q)}$. We modeled $\mathcal{E}$ as the \emph{mean absolute percentage error} (MAPE):
$\mathcal{E}(Q',\mathcal{C}\setminus\mathcal{C}') = \frac{1}{|\mathcal{C}|}\sum\limits_{c_i \in \mathcal{C}, Q'\not\models c_i}  \frac{\left\lvert \beta_i - f_{c_i}(Q') \right\rvert}{\beta_i} $.

\section{Deriving relaxations} \label{sec:heuristics}
Deriving the optimal relaxation $Q^*$ (Definition~\ref{def:best-relax}) is
computationally expensive: The space of possible relaxations is exponential in
the number of constraints of $Q$. Moreover, computing the objective value $\obj(Q')$ for each
relaxation $Q'$ is also an expensive operation, as it involves deriving the
top-1 package for $Q'$.
In this section, we propose simple but effective heuristics to avoid searching 
for $Q^*$ in the entire space of candidate relaxations. These methods are able to
find good approximations for $Q^*$ very efficiently.

\subsection{Extent of relaxation} 
\label{sub:relaxation_level}

\looseness -1
We first study how aggressively we should relax queries: Is it better to relax
more or fewer constraints? We investigate the effectiveness of relaxations when 
we exhaustively relax queries by removing a fixed amount of constraints.
Given a query $Q_{\mathcal{C}, \obj}$ and a relaxation level $k$, we
derive a relaxed query $Q^k_{\mathcal{C}^k, \obj}$, such that
$|\mathcal{C}^k|/|\mathcal{C}|=(100-k)\%$.  We use two methods to derive $Q^k$
based on Definition~\ref{def:best-relax}:
\begin{description}[noitemsep,leftmargin=0.3cm]
    \item[\exhaustiveI:] 
    maximizes the \emph{improvement} $\mathcal{I}(Q^k;Q)$, while ignoring constraint violations (i.e., $\mathcal{E}(Q^k,\mathcal{C}\setminus\mathcal{C}^k)=0$).
    
    \item[\exhaustiveIE:] 
    maximizes the \emph{improvement/error ratio},
    similarly to Equation~\eqref{eq:optimal-relax}.
\end{description}

We tested the performance of these exhaustive relaxation approaches on a
dataset of real recipes extracted from allrecipes.com. We constructed a set of
10 random package queries, in the spirit of Example~\ref{ex:query}, where
the total number of constraints ranged from 3 to 10.  Half of the queries
included a minimization objective and half included a maximization objective.

\begin{figure*}[t!b]
\centering
\begin{subfigure}[b]{0.33\textwidth}
    \centering
    \includegraphics[scale=0.26]{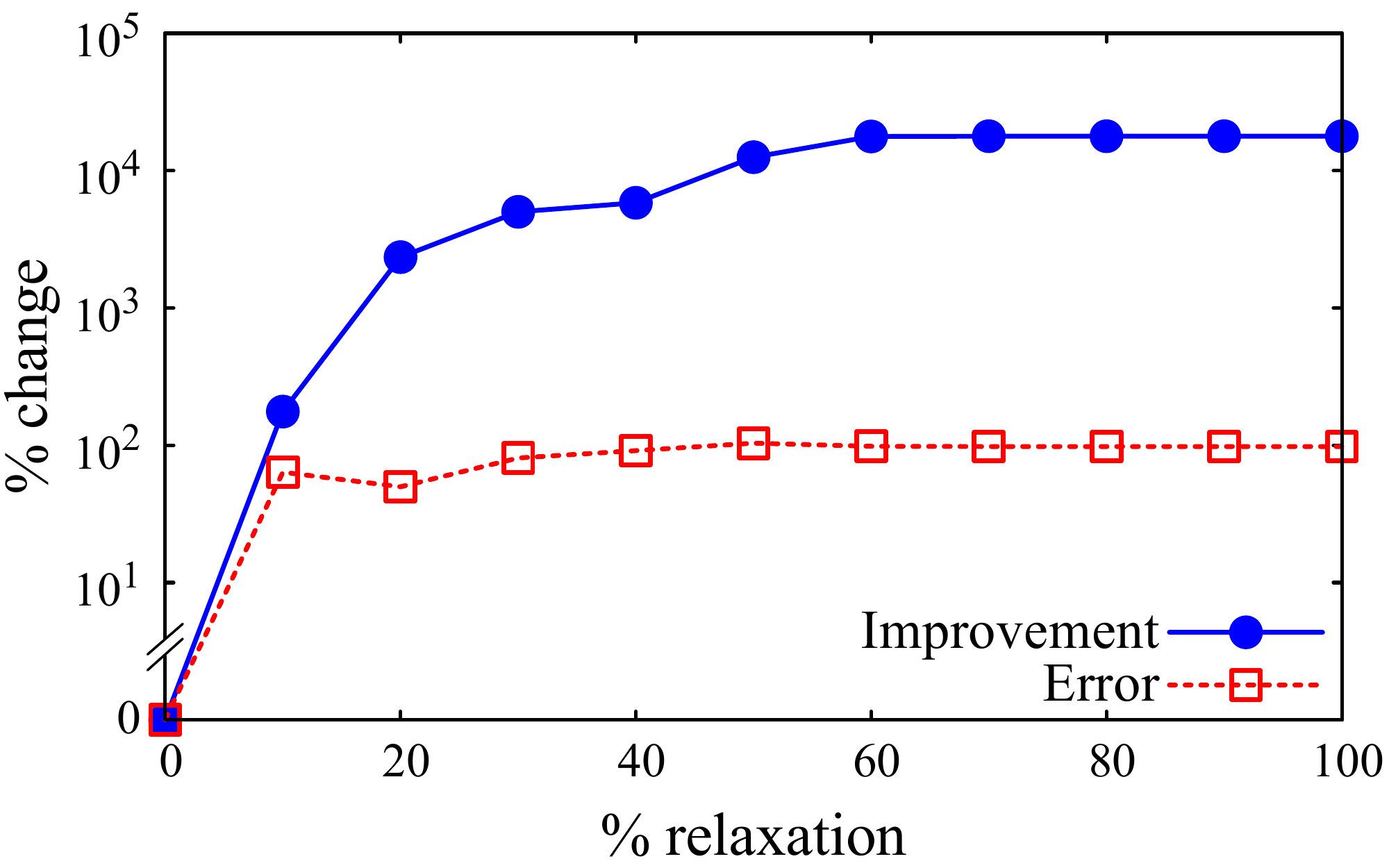}
    \caption{\exhaustiveI.}
    \label{fig:exhaustive-i}
\end{subfigure}
\begin{subfigure}[b]{0.33\textwidth}
    \centering
    \includegraphics[scale=0.26]{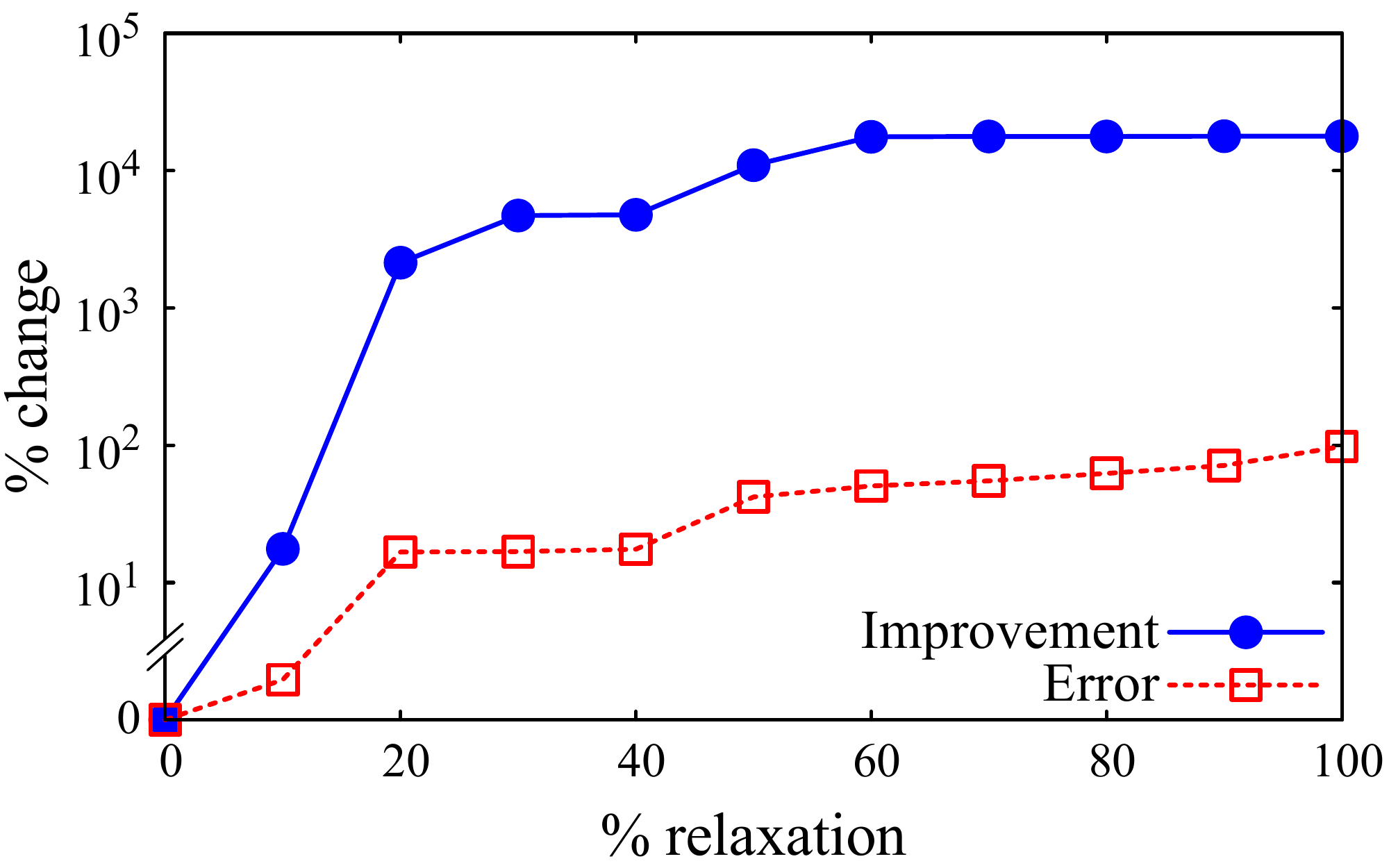}
    \caption{\exhaustiveIE.}
    \label{fig:exhaustive-ie}
\end{subfigure}
\begin{subfigure}[b]{0.33\textwidth}
    \centering
    \includegraphics[scale=0.26]{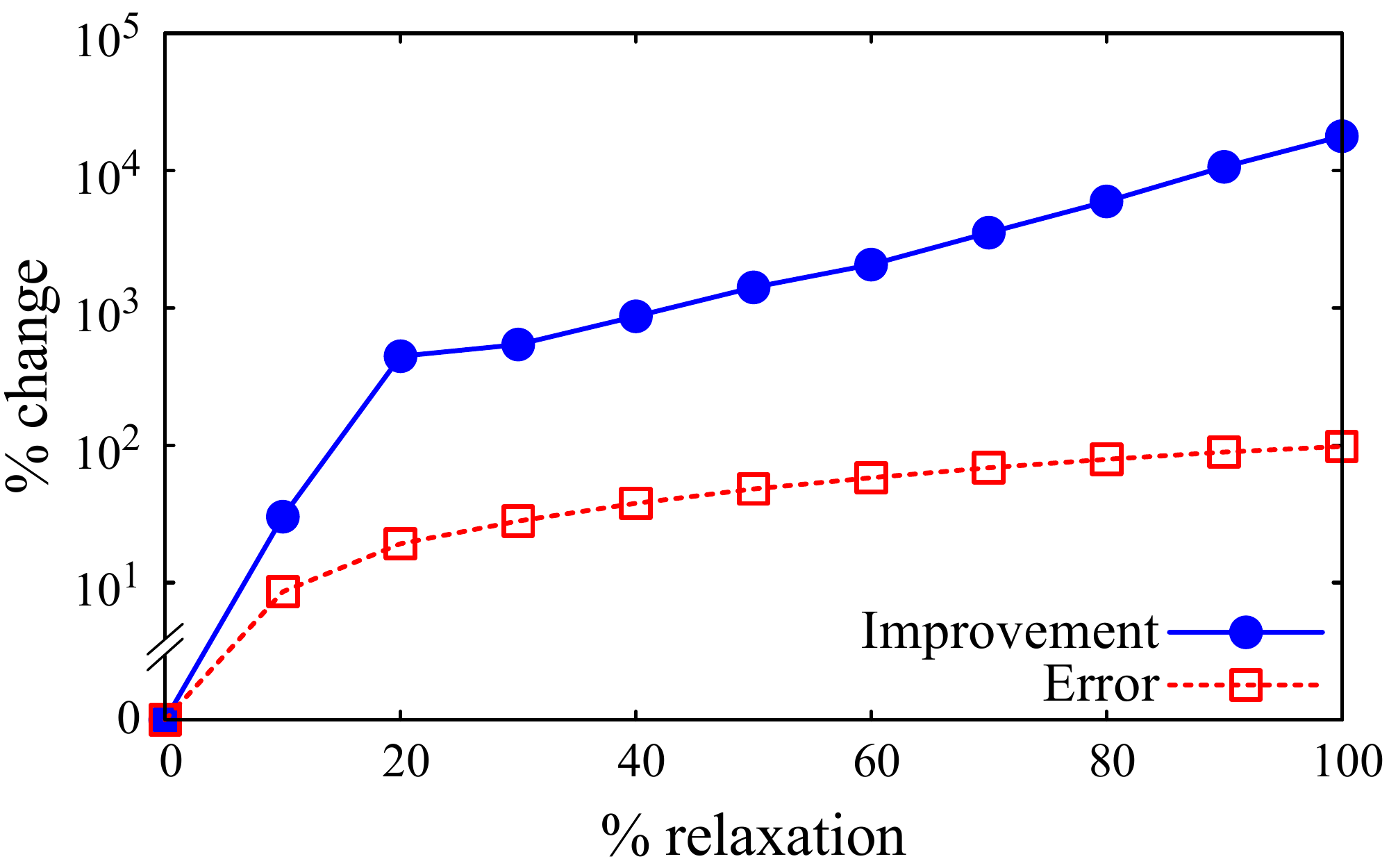}
    \caption{Random removal (no optimization).}
    \label{fig:exhaustive-random}
\end{subfigure}
\caption{Improvements and errors of exhaustive and random search. Removing more constraints increases 
both the improvement and the error. Significant improvement can be achieved after only a few 
constraints are removed. Optimizing for improvement/error ratio helps reduce the amount of 
error when less constraints are removed.}
\label{fig:exhaustive}
\end{figure*}

Figure~\ref{fig:exhaustive-i} shows the percentage improvement that
\exhaustiveI achieves in the objective function.
We see that the percentage improvement increases rapidly for small amounts of
relaxation, but the improvement slows down when we relax more aggressively. The
error line represents the mean absolute percentage error for all the
constraints that the relaxed query violates.
Following Example~\ref{ex:query}, if the
top-1 solution of a relaxed query $Q^k$ has a total of 1,000 calories, the
error due to that constraint would be $ \frac{\left\lvert 1500-1000 \right\rvert}{1500}  = 33\%$.

\exhaustiveIE behaves similarly, achieving good
improvements in the objective quickly,
but the error is generally lower, especially for
small relaxation percentages (Figure~\ref{fig:exhaustive-ie}).
For comparison purposes,
Figure~\ref{fig:exhaustive-random} shows the performance of a relaxation
algorithm that randomly chooses which constraints to relax.

This experiment demonstrates three important points: 
\begin{enumerate}[leftmargin=5mm, itemsep=-1mm, topsep=2mm]
\item The improvement curves rise sharply for low levels of relaxation, while
the gains are reduced when we relax more aggressively. This indicates that it
is
sufficient
to relax few of the constraints.

\item Removing constraints at random is worse than the
best-case approaches for low relaxation levels. This means that relaxation
cannot be arbitrary.

\item Optimizing for the improvement/error ratio can greatly reduce the error in low relaxation levels,
while achieving similar values of improvement as relaxations that ignore the
error.
\end{enumerate}

\subsection{Improving the running time}

The experiment of Figure~\ref{fig:exhaustive} shows that relaxations can achieve good improvement and error rates with the removal of a small number of constraints.  This means that focusing the relaxation search on removing one or two constraints can be an effective and practical heuristic.  In this section, we explore two iterative heuristics that greedily choose one constraint to remove at a time.
At each iteration $i$, given $Q^{i}_{\mathcal{C}^{i}, \obj}$ (initially set to $Q_{\mathcal{C}, \obj}$ at iteration $0$), each method removes:
\begin{description}[noitemsep,leftmargin=0.3cm]
\item[\greedyI:] the constraint that results in the \emph{highest improvement}: 
The algorithm selects $c_j \in \mathcal{C}^{i}$ that maximizes $I(Q^{i+1};Q^{i})$, 
where $Q^{i+1}_{\mathcal{C}^i \setminus \{c_j\}, \obj}$ is obtained from $Q^{i}$ by removing constraint $c_j$.

\item[\greedyIE:] the constraint that results in the \emph{highest improvement/error ratio}: 
The algorithm selects $c_j \in \mathcal{C}^{i}$ 
that maximizes $\frac{1 + \mathcal{I}(Q^{i+1};Q^{i})}{1 + \mathcal{E}(Q^{i+1},\mathcal{C}\setminus\mathcal{C}^{i+1})}$.
\end{description}

\begin{figure*}[t!b]
\centering
\begin{subfigure}[t]{0.33\textwidth}
    \centering
    \includegraphics[scale=0.26]{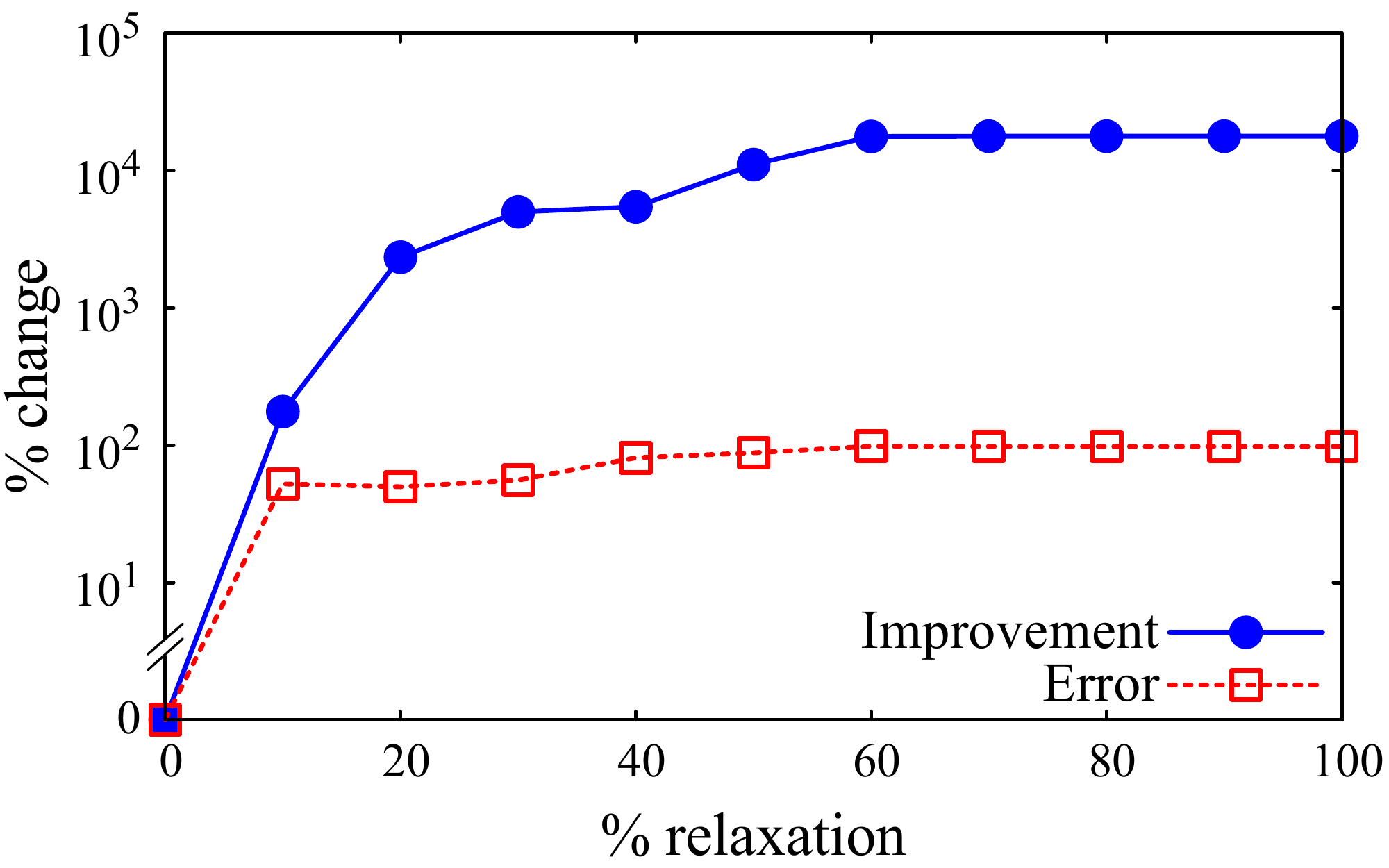}
    \caption{\greedyI.}
    \label{fig:greedy-i}
\end{subfigure}
\begin{subfigure}[t]{0.33\textwidth}
    \centering
    \includegraphics[scale=0.26]{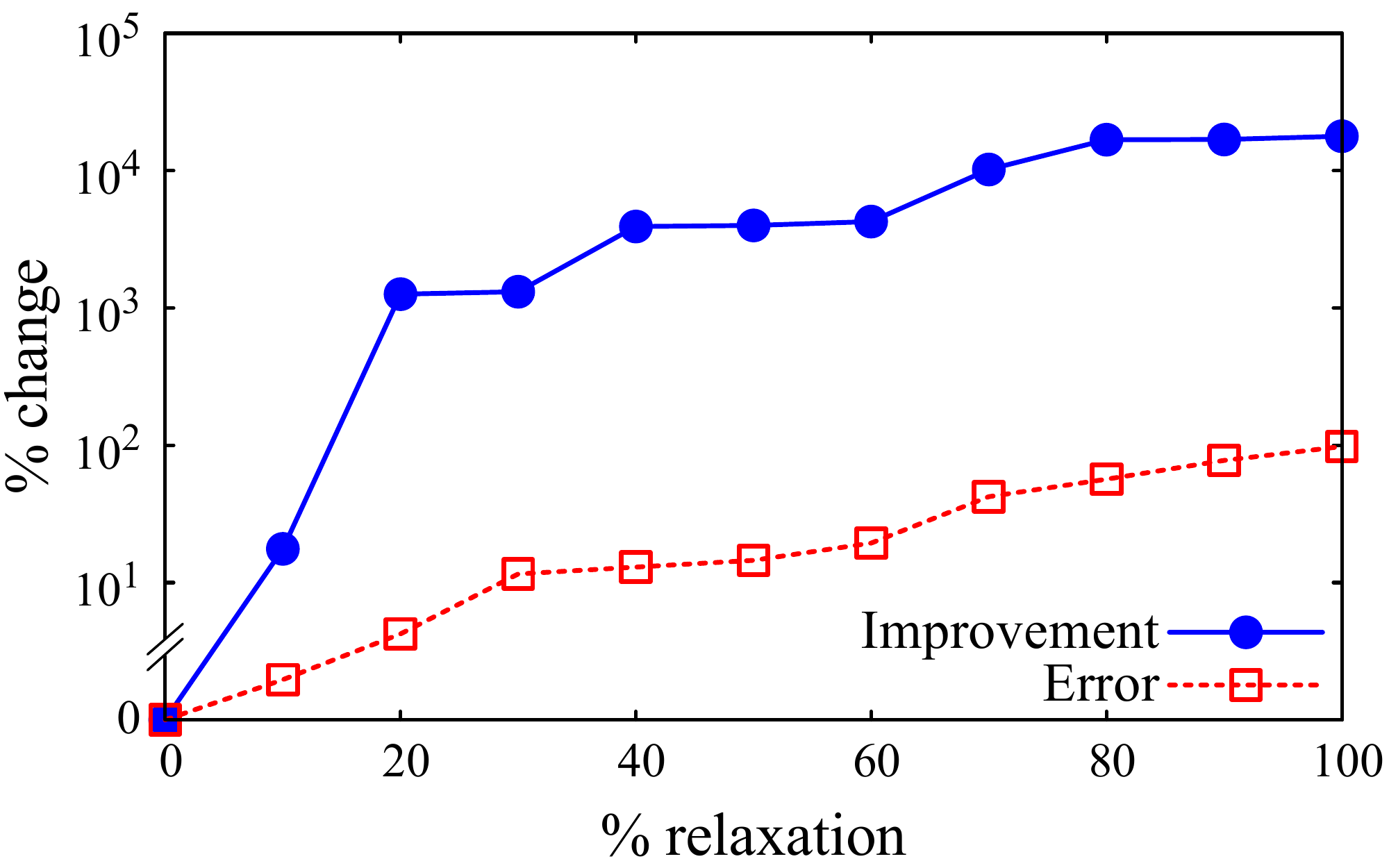}
    \caption{\greedyIE.}
    \label{fig:greedy-ie}
\end{subfigure}
\begin{subfigure}[t]{0.33\textwidth}
    \centering
    \includegraphics[scale=0.26]{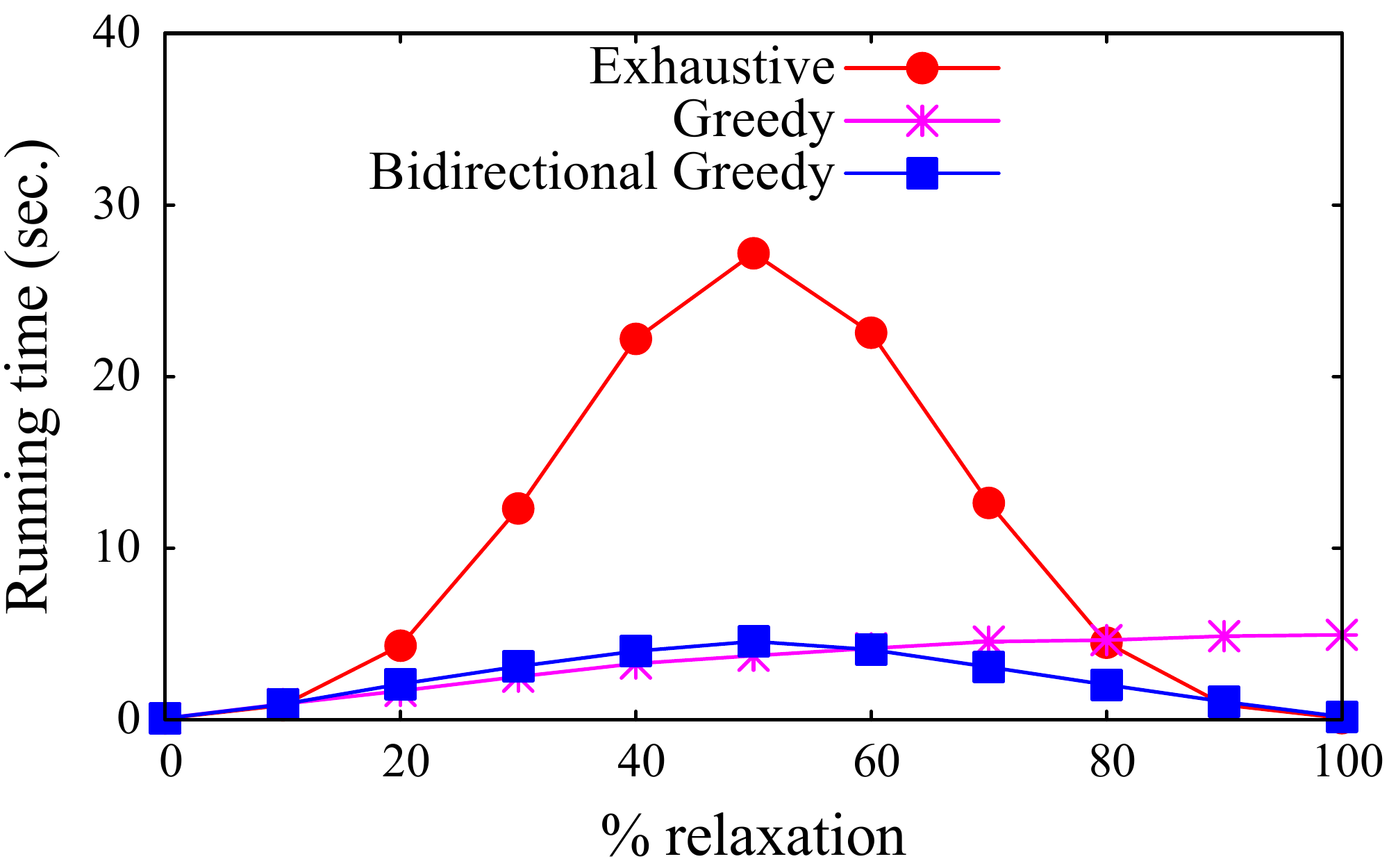}
    \caption{Running time of exhaustive and greedy search.}
    \label{fig:rtime}
\end{subfigure}
\caption{Performance of the greedy heuristics. Both versions of greedy search can achieve high improvements and low errors similarly to exhaustive search, while significantly reducing the running time.}
\label{fig:greedy}
\end{figure*}

\begin{description}[itemsep=1mm,leftmargin=0cm]
\item[Effectiveness.]
We evaluated the greedy heuristics on the same dataset and queries as the
previous experiment. The results are presented in Figure~\ref{fig:greedy}. The
two heuristics achieve similar improvement and error as their exhaustive
counterparts, especially for low levels of relaxation
(Figures~\ref{fig:greedy-i} and \ref{fig:greedy-ie}).

\item[Efficiency.]
As expected, the greedy heuristics perform much better in terms of running
time (Figure~\ref{fig:rtime}). There is no runtime difference between
\greedyI and \greedyIE, thus their performance is indicated with a single line
``Greedy'' in Figure~\ref{fig:rtime}. We improve the efficiency of the greedy heuristics even further by
introducing \emph{bidirectional greedy} as an optimization: Bidirectional
greedy starts from the non-relaxed query and removes one constraint at a time
when the target relaxation level is below 50\%. When the target relaxation
level is above 50\%, the heuristic starts from the empty constraint set and
adds one constraint at a time, until it reaches the desired level. This
optimization achieves further runtime improvements, keeping the runtime of
bidirectional greedy 
always below exhaustive search.

\end{description}

\section{User-aware relaxation} \label{sec:study}

In this section, we evaluate the usability of our approach with a crowdsourced user-study. In the study, we ask crowd workers to evaluate the quality of different
meal-plan package recommendations with respect to a set of given nutritional
requirements and other specifications. With this study, we aim to answer two
research questions:\looseness -1
\begin{description}[noitemsep,leftmargin=0.3cm]
    \item[RQ1:] Are users willing to accept relaxed recommendations?
    \item[RQ2:] Do users have preferences with respect to the types of constraints to be removed?
\end{description}

We start by describing the dataset we used, the crowd task design, and
the characteristics of the data we collected. We then proceed to discuss our
results across these two fronts.

\subsection{Experiment setup}

We evaluate the quality of package recommendations for a meal planning
application. 

\paragraph*{Dataset} \label{dataset}
We built a meal planner recommendation system with real recipe data
collected from \href{allrecipes.com}{allrecipes.com}. We collected 
7,955 recipes with information on ingredients, nutritional content, and
preparation time. Our application constructs meal plan recommendations
given constraints on preparation time and recipe contents.

\begin{figure*}[ht!]
    \centering
        \includegraphics[width=\textwidth]{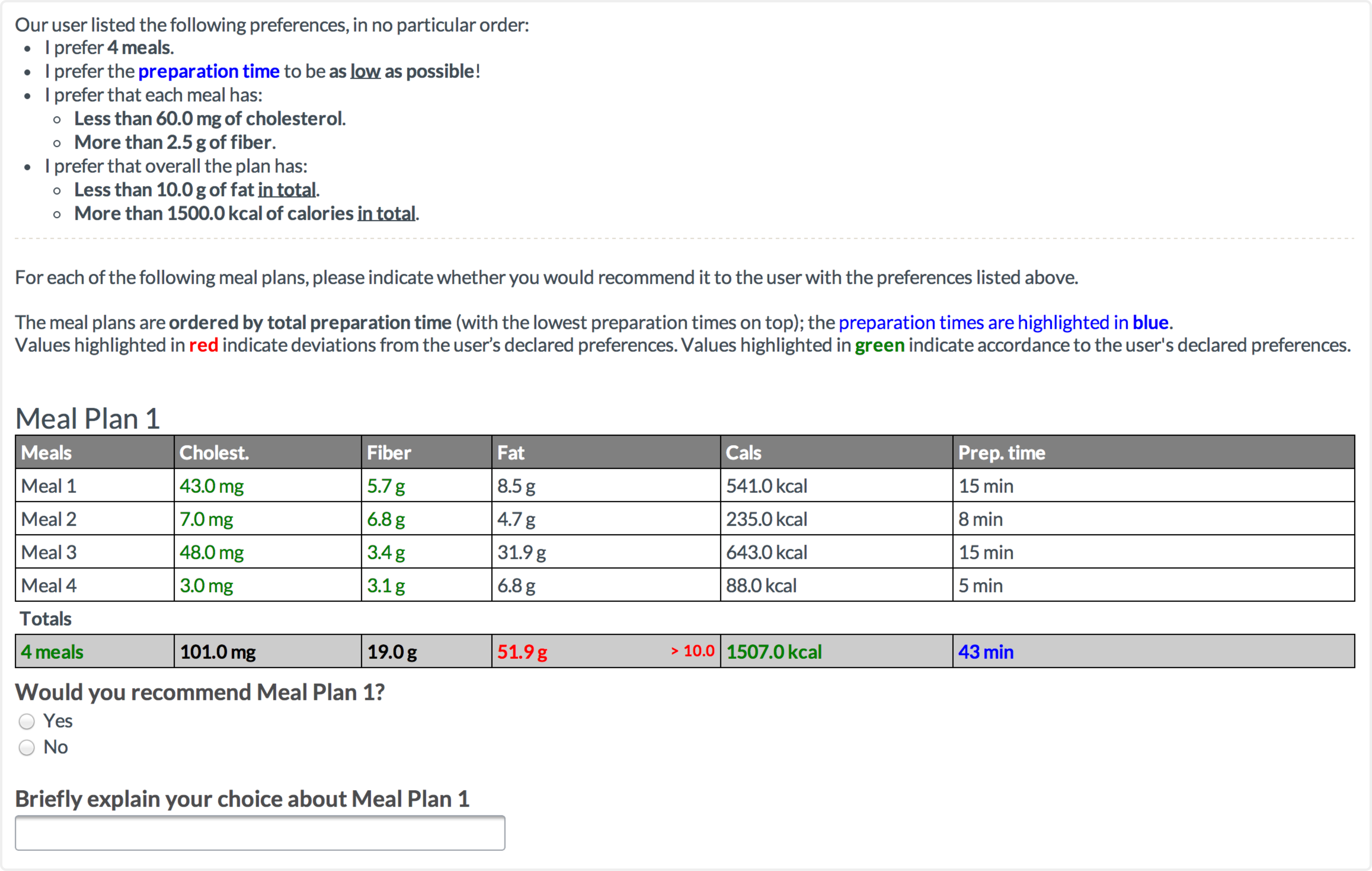}
        \vspace{-5mm}
    \caption{In our crowd task, we provide workers with a list of user
    preferences and we ask them to choose whether to recommend each one of
    five meal plans. The five meal plans listed in the task always include one
    plan that satisfies the user constraints, three plans that relax one
    constraint each, and one random plan.}
    \label{fig:figures_taskScreenshot}
\end{figure*}

\paragraph*{Task design}
We designed a crowdsourcing task to evaluate (a) whether users like
recommendations that relax one of the package constraints, and (b) whether
they are sensitive to some constraints more than others. In the task scenario,
we tell the crowd workers that we need to recommend some meal plans to a user,
given a set of preferences that the user specified. The user specifications
include four types of preferences:
\begin{description}[leftmargin=5mm, itemsep=-1mm]
    \item[Cardinality constraint:] The number of meals in the plan.
    \item[Objective criterion:] All task instances express that
    the user prefers the preparation time for the meal plan to be as low as
    possible.
    \item[Base constraints:] Two constraints on the nutritional content of
    each meal in the plan.
    \item[Global constraints:] Two constraints on the nutritional content of
    the entire meal plan.
\end{description}

After listing the user preferences, the task displays five alternative meal
plans and asks workers to select, for each plan, whether they would recommend
it to the user.  
Each task includes the following five meal plan types:
\begin{description}[leftmargin=5mm, itemsep=-1mm]
    \item[\original:] It strictly adheres to the constraints specified in the user preferences.
    \item[\cardinality:] It relaxes the cardinality constraint.
    \item[\base:] It relaxes one of the base constraints.
    \item[\globalRelax:] It relaxes one of the global constraints.
    \item[\random:] It is a collection of randomly selected meals that satisfies the cardinality constraint.
\end{description}

The plans are computed automatically, using the \textsc{PackageBuilder}
engine~\cite{pbq-demo}, and are listed in increasing order of preparation
time. We use colors to indicate adherence to or violation of the constraints,
which helps the workers easily note which constraints were violated. 
Figure~\ref{fig:figures_taskScreenshot} shows a partial screenshot of
the task. We chose to omit recipe names to avoid worker bias due to personal
preferences (e.g., biases due to religious dietary restrictions or cuisine).

\paragraph*{Collected data}
We automatically generated 50 different unique configurations of our task
on the Crowdflower
platform.\footnote{\href{http://www.crowdflower.com/}{http://www.crowdflower.com/}}
Each
configuration was completed by 10 unique workers, and each worker was not
allowed to complete more than 5 configurations.

We used the explanation field to identify and remove obvious spammers from the
dataset. We rejected workers who gave the same answer and comment in every
task they completed, workers who entered random data in the explanation field,
and workers who gave explanations that were inconsistent with their
selections. After cleaning, our dataset included 115 unique workers, and 306
unique task instances.

Figure~\ref{fig:study-results} summarizes our results. The first table shows
the number of times that each type of plan was recommended.
We note that in 16 cases, users recommended the \random plan, which could mean
that the dataset may still contain a few spammers that we were not able to
identify. 
The other two
tables show the number of times that each relaxation result was recommended in
the cases that the \original plan was accepted or rejected, respectively.

\begin{figure*}[t!b]
    \begin{minipage}[t]{0.3\linewidth}
        {\small
        \begin{tabular}{lrr}
                \multicolumn{3}{l}{\textbf{Overall}}\\
                \toprule
                \textbf{method}  & \multicolumn{2}{c}{\textbf{recommended}}\\
                \midrule
                \original   &    216 &(70.59\%)\\
                \base       &    182 &(59.48\%)\\
                \globalRelax&    94  &(30.72\%)\\
                \cardinality&    76  &(24.84\%)\\
                \random     &    16  &(5.23\%)\\
                \bottomrule
        \end{tabular}}
    \end{minipage}
    \hfill
    \begin{minipage}[t]{0.3\linewidth}
        {\small
        \begin{tabular}{lrr}
                \multicolumn{3}{l}{\textbf{When \original is recommended}}\\
                \toprule
                \textbf{method}  & \multicolumn{2}{c}{\textbf{recommended}}\\
                \midrule
                any         &    150 &(69.44\%)\\
                \base       &    118 &(54.63\%)\\
                \globalRelax&    63  &(29.17\%)\\
                \cardinality&    47  &(21.76\%)\\
                \bottomrule
                \\
        \end{tabular} }
    \end{minipage}
    \hfill
    \begin{minipage}[t]{0.3\linewidth}
        {\small
        \begin{tabular}{lrr}
                \multicolumn{3}{l}{\textbf{When \original is rejected}}\\
                \toprule
                \textbf{method}  & \multicolumn{2}{c}{\textbf{recommended}}\\
                \midrule
                any         &    82  &(91.11\%)\\
                \base       &    64  &(71.11\%)\\
                \globalRelax&    31  &(34.44\%)\\
                \cardinality&    29  &(32.22\%)\\
                \bottomrule
                \\
        \end{tabular} }
    \end{minipage}
    \caption{Our study showed that relaxations may often be preferred to the
    non-relaxed solutions. Even when the \original plan is chosen, users
    still recommend at least one relaxation almost 70\% of the time. When
    \original is not chosen, users find an acceptable result among the relaxed
    plans more than 90\% of the time.}\label{fig:study-results}
    \vspace{-2mm}
\end{figure*}

\subsection{RQ1: Evaluation of relaxations}

We first analyze our dataset to evaluate whether relaxations are useful.
Our data shows that users are often dissatisfied with the meal plan that
follows all of the specified constraints. From the 306 completed tasks in our
dataset, the \original plan is rejected about 30\% of the time. The reason is
usually that users consider the value of the objective criterion
unacceptable. This means that relaxations are needed, even when the
original recommendation set is not empty.

We further observe that users are in fact \emph{more likely to choose a
relaxed plan}, than the plan that strictly adheres to the specified
constraints. Out of the 306 tasks, users selected at least one of the relaxed
plans (\base, \globalRelax, or \cardinality) 75.82\% of the time, which is more
than the number of times they recommended the \original plan.

Finally, our study shows that even when the \original plan is selected, the
users also recommend one or more of the relaxations in almost 70\% of the
cases. More importantly, in the many cases where the \original plan is
rejected, the users find an acceptable plan within the relaxations in more
than 90\% of the cases.

\vspace{1mm}
\noindent
\textbf{Conclusion:} Query relaxation is a powerful technique for package
recommendations, and often produces packages that are preferable to the
non-relaxed solutions.

\subsection{RQ2: Constraint sensitivity} 
\label{sub:constraint_sensitivity}
When considering query relaxations, it is important to know whether a
relaxation algorithm should prioritize certain constraints. In our experiment,
our 115 unique users showed a clear preference for relaxations of base
constraints (\base was chosen in almost 60\% of cases), and they were least
likely to prefer relaxations of cardinality constraints (about 25\% of
cases). We observe similar behaviors when the \original plan is either accepted or
rejected.

We do not believe, however, that this finding generalizes:
Depending on the application and dataset, we may observe different behaviors.
For example, a user
with celiac disease is unlikely to relax a base constrain on gluten. 

\vspace{1mm}
\noindent
\textbf{Conclusion:} The important takeaway of these observations is that, for
a given application and dataset, users can strongly prefer one type of relaxation over
another, and relaxation techniques should exploit this. These preferences may
not always be known a priori, but they may be derived from past use-data of the
system.

\subsection{Additional lessons and discussion}\label{sec:discuss} 

We manually examined the explanations entered by workers in order to understand
the reasoning for selecting or rejecting a particular plan. The workers often
cited lower preparation time as the reason for selecting a
relaxation. For example, one user accepted a package because it was a ``\emph{close match for fiber as
required and less time}.'' Another user notes: ``\emph{Even thought [sic] the protein is low this
is the best with a low prep time}.''
This verifies our intuition that a good recommendation should improve some
aspect of the original query (e.g., the objective criterion).

Our users often incorporated application-specific logic in their selections.
For example, one person selected a \base plan because ``\emph{adding
some proteins to the meal could be easy.}'' This shows that relaxation
algorithms can use application-specific knowledge to determine constraint
priorities.

Finally, some user comments provided a good explanation of the strong bias that we observed toward relaxations of base constraints: 

\smallskip
\noindent
``\emph{Since your preference is 60 mg of cholesterol per meal the overall will be 240 mg, so its okay.}''\footnote{Recall that, in the task, preference was given to plans with 4 meals.}

\smallskip
\noindent
``\emph{This meal plan meets most preferences. Two of the meals are lower in protein but two are high in protein which balances it out.}''
\smallskip

\looseness -1
In these cases, workers applied a type of relaxation that we did not provide
to them: They transformed a base constraint into a global constraint. This relaxation can be applied to base constraints on numeric values. We will explore this type of
relaxation in future work.

\section{Related Work}\label{sec:related}

Research in relational query modification (relaxation or refinement) attempts to mitigate two of the problems of the precise database answer model: (i) the empty-answer problem, or (ii) the too-many-answers problem ~\cite{preference-survey, joinrelax-koudas, querymodification-surajit}. Stefanidis et al. survey several techniques for encoding user preferences as soft constraints to improve the query results~\cite{preference-survey}.

In contrast to its traditional use, we use query relaxation to
\emph{improve} the results of queries that do produce enough results, but whose ``quality''
is limited by over- or poorly- specified constraints. In such cases, queries can be slightly relaxed
to allow the exploration of a larger pool of feasible solutions and, hopefully, the discovery
of more interesting solutions.

Package recommendation systems have only recently received attention from the research community.
They have been used to derive \emph{travel plans}~\cite{xie}, \emph{team formations}~\cite{Lappas2009,Anagnostopoulos:2010},
\emph{course combinations}~\cite{course-rank,Parameswaran2010}, \emph{composite items}~\cite{BasuRoy:2010},
as well as nutritionally balanced \emph{meal plans}~\cite{pbq-demo}. 
Package recommendations are also close to \emph{preference queries over sets}~\cite{zhang2011preference} and skyline groups~\cite{Li:2012}, which additionally involve multiple objective criteria.
Recent work by Deng et al. discusses the complexity of computing package recommendations and the complexity of searching for query relaxations~\cite{deng}.

\section{Contributions and next steps}\label{sec:directions}
In this paper, we proposed a novel method for deriving package recommendations
based on query relaxation. Our crowd user study showed that users often prefer
relaxed solutions to non-relaxed results, which indicates that this
is a very promising direction for handling package recommendations.

Future research will extend our work in the following directions:
\begin{itemize}[leftmargin=5mm, topsep=1mm]
    \item Our current user study showed a clear bias toward relaxing base
    constraints. We plan to conduct a larger study to investigate what dictates
    user sensitivity toward different kinds of constraints.
    
    \pagebreak
    \item In this work, we focused on coarse relaxation (removing constraints) to test the viability of query relaxation as a package recommendation method.  Even at this granularity, relaxation showed great promise, both on our effectiveness experiments and the user study.  We believe that finer granularity relaxations will give us better control to cater to user preferences.
    
    \item We will explore additional relaxation methods including the one that our crowd users produced unprompted: relaxing base constraints into global constraints.
    
    \item We will study the design of more efficient relaxation algorithms that take into account knowledge of data distributions.
\end{itemize}

\smallskip
\noindent
\textbf{Acknowledgements.}
This paper was presented at the Data4U workshop at PVLDB 2014. This
work was partially supported by the National Science Foundation under grants
IIS-1421322 and IIS-1420941.

\bibliographystyle{abbrv}
\bibliography{refs}

\end{document}